# Crossover of ballistic, hydrodynamic, and diffusive phonon transport in suspended graphene

Xun Li[1] and Sangyeop Lee[1,2*]

[1] *Department of Mechanical Engineering and Materials Science,*
*University of Pittsburgh, Pittsburgh PA 15261*

[2] *Department of Physics and Astronomy,*
*University of Pittsburgh, Pittsburgh PA 15261*

*sylee@pitt.edu

**Abstract**

Hydrodynamic phonon transport was recently predicted as an important regime for phonon transport in graphitic materials. Many of past studies on hydrodynamic phonon transport have focused on the cases where the hydrodynamic regime is significantly stronger than other regimes such that hydrodynamic features can be clearly observed. However, this often requires stringent conditions of temperature and sample size. In many cases, the transport cannot be characterized by a single regime, but the features of all three regimes – ballistic, hydrodynamic, and diffusive regimes – exist to some extent. Here we assess the extent of three regimes by comparing momentum destruction rates by three different mechanisms, each of which represents a different regime: diffuse boundary scattering without internal phonon scattering (ballistic regime), diffuse boundary scattering combined with normal scattering (hydrodynamic regime), and umklapp scattering (diffusive regime). We solve the Peierls-Boltzmann equation with an *ab initio* full scattering matrix using a deviational Monte Carlo method. We sample distribution functions of ballistic and scattered particles separately, and thereby compare the momentum destruction rates by the three different mechanisms. Using this framework, we discuss a well-known phenomenon of ballistic-to-hydrodynamic crossover, called phonon Knudsen minimum.

## 1. Introduction

With rapidly increasing demand for efficient heat removal from electronic devices[1], graphene has drawn significant interests due to its extremely high thermal conductivity[2], [3] as well as its high charge mobility[4]. Phonon transport in suspended graphene is known to deviate from the diffusive regime where the Fourier's law of heat conduction is valid, and this deviation can be associated with weak resistive umklapp scattering (hereafter U-scattering)[5]. It was found through first-principle-based calculations that normal scattering (hereafter N-scattering), which conserves total crystal momentum of phonons (hereafter phonon momentum), is the primary mechanism of phonon scattering in suspended graphene[5], [6]. Different from the diffusive regime, the regime where N-scattering dominates other types of scattering processes is called hydrodynamic regime. The hydrodynamic phonon transport was recently predicted to be significant in graphitic materials including suspended graphene[7], [8], single-wall carbon nanotubes[9], and graphite[10].

Many of past studies on the hydrodynamic phonon transport have focused on the cases where the hydrodynamic features are significantly strong. The strong hydrodynamic regime was discussed using two representative phenomena: phonon Poiseuille flow[7], [8], [11]–[16] and second sound[7]–[9], [11], [17], [18]. The former is the steady-state phonon flow where thermal resistance is due to diffuse boundary scattering combined with N-scattering (i.e., hydrodynamic viscous damping[15]), and the latter is the propagation of a temperature wave without significant damping. The existence of phonon Poiseuille flow could be confirmed through the temperature and sample width dependences of thermal conductivity that increases faster than that in the ballistic regime[7,12]. Recent studies showed these peculiar dependences, by solving the Peierls-Boltzmann equation (PBE) either with the Callaway's scattering model[7], [10], [14] or with a full scattering matrix[15]. The unique dependences are the result of the diffuse boundary scattering combined with N-scattering, which causes less thermal resistance compared to the case where there is no internal phonon scattering. This effect was interpreted as friction induced on the relaxon gas[19] or hydrodynamic viscous damping[15]. The hydrodynamic viscous damping is the major thermal resistance mechanism in the hydrodynamic regime and can explain the peculiar dependences of thermal conductivity[15].

However, in many practical situations when the conditions for the strong hydrodynamic regime are too stringent to be met, the transport is in the crossover of different regimes. In these cases, the features from all three regimes – the ballistic, hydrodynamic, and diffusive regimes – can coexist. The transport phenomena then become complicated and not easy to understand. A map of three transport regimes in the temperature and sample size space was often determined by comparing mode-averaged mean free path (MFP) with sample size[7], [8], [11], [20]. For example, Guyer and Krumhansl suggested a condition for the phonon Poiseuille flow as

$$\lambda_{\mathrm{N}} \quad R \quad \sqrt{\lambda_{\mathrm{N}}\lambda_{\mathrm{U}}} \tag{1}$$

where $\lambda_{\mathrm{N}}$ and $\lambda_{\mathrm{U}}$ are mode-averaged MFPs of N- and U-scattering[11]. The $R$ is a characteristic sample size, e.g., the diameter of a rod for three-dimensional materials. If $R$ is much smaller than $\lambda_{\mathrm{N}}$, the transport is ballistic; if $R$ is much larger than $\sqrt{\lambda_{\mathrm{N}}\lambda_{\mathrm{U}}}$, the transport is diffusive. However, the MFPs of N- and U-scattering often have an extensively wide range with respect to phonon frequency, causing an overlap with each other in some cases. In addition, the criteria are useful only when the characteristics of one regime are significantly stronger than those from other regimes, and cannot properly describe a transition across two different regimes. The three parameters ($\lambda_{\mathrm{N}}$, $R$, and $\sqrt{\lambda_{\mathrm{N}}\lambda_{\mathrm{U}}}$) are often in the similar order, and thus the transport regime has the characteristics of all three regimes in many cases. Then, the transport phenomena cannot be described by a single transport regime, and the detailed understanding of such cases is still lacking.

For the cases where the transport features from all three regimes coexist, it is desired that we can quantitatively measure the extent of all three regimes. For this purpose we can check the momentum balance of a phonon system. Regardless of its transport regime, phonons are driven by a temperature gradient. In other words, the phonon system gains a net momentum from a temperature gradient. When the phonon flow is at steady state, the momentum gain is balanced with momentum destruction by three mechanisms, each of which represents a different regime: the diffuse boundary scattering without internal phonon scattering (ballistic regime), the diffuse boundary scattering combined with N-scattering (hydrodynamic regime), and U-scattering (diffusive regime). Analyzing the momentum balance is directly relevant to understanding thermal transport phenomena when the hydrodynamic regime contributes to the actual phonon transport. In an ideal hydrodynamic regime where phonons exhibit a collective motion, the heat flux is linearly proportional to the net momentum of the phonon system.

In this paper, we quantitatively measure the momentum destruction rates by the aforementioned three mechanisms in suspended graphene. We solve the PBE with an *ab initio* full scattering matrix using a deviational Monte Carlo (MC) method[15], [21]. An advantage of MC simulation is that we can sample distribution functions for scattered and unscattered particles separately[22]. By doing so, we can quantitatively show the contributions of thermal conductivity and momentum destruction rate from the three regimes.

## 2. Method and approach

We solve the PBE for infinitely long suspended graphene with finite width shown in Fig. 1. We use the deviational MC method to solve the energy-based PBE with a full scattering matrix[15], [21]. The deviational MC method for the single mode relaxation time (SMRT) approximation has been validated against analytic solution in previous studies[21], [23]. The validation of MC method employing a full scattering matrix is provided in the Supplementary Information[24]. We compare the thermal conductivity values of an infinitely large graphene sample from our MC method and full-iterative method that has been widely used in recent studies[5], [25]–[30]. The phonon dispersion and full scattering matrix are obtained from lattice dynamic calculations using second and third order force constants, which are from density functional theory. The details of calculation can be found somewhere else[25], [31]. At steady state, the energy-based deviational PBE with a linearized scattering operator is

$$\mathbf{v}_{\mathbf{q}s} \cdot \nabla_{\mathbf{r}} f^{\mathrm{d}}_{\mathbf{q}s} = \sum_{\mathbf{q}'s'} B_{\mathbf{q}s,\mathbf{q}'s'} f^{\mathrm{d}}_{\mathbf{q}'s'} \tag{2}$$

where $f^{\mathrm{d}}_{\mathbf{q}s}$ is the deviational energy distribution function at the phonon state with wavevector $\mathbf{q}$ and polarization $s$. The $f^{\mathrm{d}}_{\mathbf{q}s}$ is defined as $\omega\left(n-n^{0}\right)$ where $\omega$, $n$, and $n^{0}$ are phonon energy, phonon distribution function, and the Bose-Einstein distribution at equilibrium temperature, respectively. The left side of Eq. (2) describes the advection of phonons where $\mathbf{v}_{\mathbf{q}s}$ is a phonon group velocity and $\mathbf{r}$ is a spatial vector. The right-hand side is the change of phonon energy distribution due to internal phonon scattering. The element $B_{\mathbf{q}s,\mathbf{q}'s'}$ is the scattering matrix in the energy-based PBE, defined as $C_{\mathbf{q}s,\mathbf{q}'s'}\omega_{\mathbf{q}s} / \omega_{\mathbf{q}'s'}$, where $\mathbf{C}$ is the scattering matrix in the PBE. Our *ab initio* full scattering matrix includes three-phonon scattering process only.

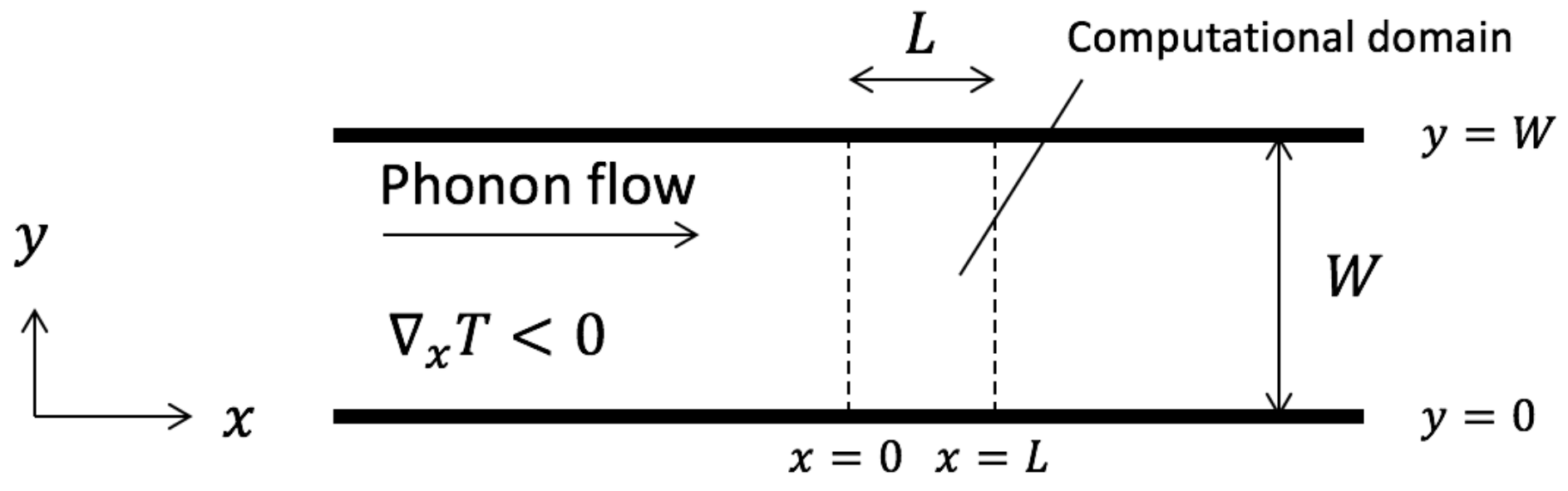

Fig. 1. A schematic picture of an infinitely long graphene sample for MC simulation.

The advection and scattering of phonon particles are explicitly simulated by allowing particles to travel with their group velocities, performing internal phonon scattering based on the scattering matrix, and applying diffuse boundary scattering if they cross the boundaries. For internal phonon scattering, we follow the previous work[21] and introduce a matrix $\mathbf{P}$ calculated from the *ab initio* scattering matrix:

$$\mathbf{P}(\Delta t) = e^{\mathbf{B}\Delta t} = \sum_{k=0}^{\infty} \frac{\Delta t^k}{k!} \mathbf{B}^k \tag{3}$$

where the matrix $\mathbf{P}$ is defined such that

$$f_i^{\mathrm{d}}(t+\Delta t) = \sum_j P_{ij}(\Delta t) f_j^{\mathrm{d}}(t) \tag{4}$$

The phonon state indices $\mathbf{q}s$ and $\mathbf{q}'s'$ are replaced with $i$ and $j$ for the sake of simplicity. The matrix $\mathbf{P}$ correlates the distribution functions at $t$ and $t+\Delta t$. A stochastic algorithm of phonon scattering using $\mathbf{P}$ can be found in literatures[15], [21].

In this work, we divide phonon particles into scattered and ballistic particles depending on whether a particle has experienced internal phonon scattering (scattered) or not (ballistic). In the MC simulation, each particle carries a label indicating ballistic or scattered. All particles generated from initialization, temperature gradient, or diffuse boundary scattering are labeled as ballistic particles as they did not experience internal phonon scattering. The diagonal terms of the matrix $\mathbf{P}$ represent the case where neither N- nor U-scattering occurs during $\Delta t$ and thus the phonon state is not changed. Therefore, upon an internal phonon scattering, the distribution function for the ballistic particles can be updated with

$$f_{i,\mathrm{ballistic}}^{\mathrm{d}}(t+\Delta t) = P_{ii}(\Delta t) f_{i,\mathrm{ballistic}}^{\mathrm{d}}(t) \tag{5}$$

which describes a ballistic particle at $t$ remains as ballistic at $t+\Delta t$. The distribution function of scattered particles can be updated with

$$f_{i,\text{scattered}}^{\text{d}}\left(t+\Delta t\right)=\sum_{j\neq i}P_{ij}\left(\Delta t\right)f_{j,\text{ballistic}}^{\text{d}}\left(t\right)+\sum_{j}P_{ij}\left(\Delta t\right)f_{j,\text{scattered}}^{\text{d}}\left(t\right) \tag{6}$$

The first term includes off-diagonal terms of $\mathbf{P}$ representing scattering. The second term is the contribution from scattered particles at $t$ regardless of whether those particles experienced scattering or not during the time from $t$ to $t+\Delta t$. Simply speaking, if a particle that is labeled as ballistic does not change its phonon state after $\Delta t$, it remains as ballistic. If a particle is determined to change its state, it is labeled as scattered. The ballistic and scattered distributions are calculated at each time step, by counting the number of particles for each label.

The local heat flux at $\mathbf{x}$ from the ballistic and scattered particles can be expressed as

$$q''_{\alpha,x}(\mathbf{x})=\frac{1}{N_i V_{\text{uc}}}\sum_i v_{i,x}\ \ \omega_i n_{\alpha,i}(\mathbf{x}) \tag{7}$$

where $\alpha$ indicates ballistic or scattered label. The $N_i$ and $V_{\text{uc}}$ are the number of phonon states and volume of the unit cell. In our MC simulation, the local heat flux can be calculated by summing the heat fluxes of all particles in a small grid volume. Then, the local thermal conductivity, $\kappa_{\alpha,xx}(\mathbf{x})$, can be simply found as $-q''_{\alpha,x}(\mathbf{x})/\nabla_x T$.

For the momentum balance analysis, we would need to calculate the wall shear stress, which can be defined as the rate of momentum destruction by the diffuse boundary scattering per given time and area. The wall shear stress at the bottom wall ($y=0$) can be expressed as

$$\tau_\alpha(y=0)=\frac{1}{N_i V_{\text{uc}}}\sum_{i,v_{i,y}<0} q_{i,x}v_{i,y}n_{\alpha,i}(y=0) \tag{8}$$

The wall shear stress in our MC simulation can be calculated by summing the $x$-direction momentum of particles that cross the boundary during a given time interval.

Fig. 1 shows a schematic of sample geometry and thermal gradient direction. The sample is infinitely long in $x$-direction. In this case, the phonon distribution function is invariant along the $x$-direction except for the change due to a temperature gradient. Therefore, our boundary conditions at $x=0$ and $x=L$ are[32]

$$f(x=0)=f(x=L)+\frac{df^0}{dT}L\nabla_x T \tag{9}$$

where $f^0$ is the energy Bose-Einstein distribution function, $\omega n^0$. For the top and bottom boundaries at $y = 0$ and $y = W$, an adiabatic boundary condition with complete diffuse boundary scattering is applied. The time, real, and reciprocal space domains are discretized. The time step is chosen such that it is smaller than the minimum lifetime of the phonons with frequencies below $k_{\mathrm{B}}T /$ . The real-space domain is discretized uniformly into 20 control volumes along $y$-direction and one control volume with a length of 10 nm along $x$-direction where a temperature gradient of 1000 K/m is applied. A 40×40 grid is used to sample the reciprocal space. It was confirmed that the calculation results reasonably converge with respect to all discretization variables.

## 3. Results and discussion

### 3.1. Decomposition of thermal conductivity

In Fig. 2, we show the total and decomposed thermal conductivity values with respect to sample width at 100 K. Both contributions from the ballistic and scattered particles increase with increasing sample width, but the dominant contributor differs for different sample width. At small width below 100 nm, the ballistic contribution is much larger than the scattered contribution since internal phonon scattering is very weak compared to the diffuse boundary scattering. In the mid-range of sample width, the ballistic and scattered contributions have a crossover, indicating that the dominant transport regime is changed from ballistic to non-ballistic regimes. As width increases further and becomes larger than 1 μm, most of the heat flux comes from the scattered particles.

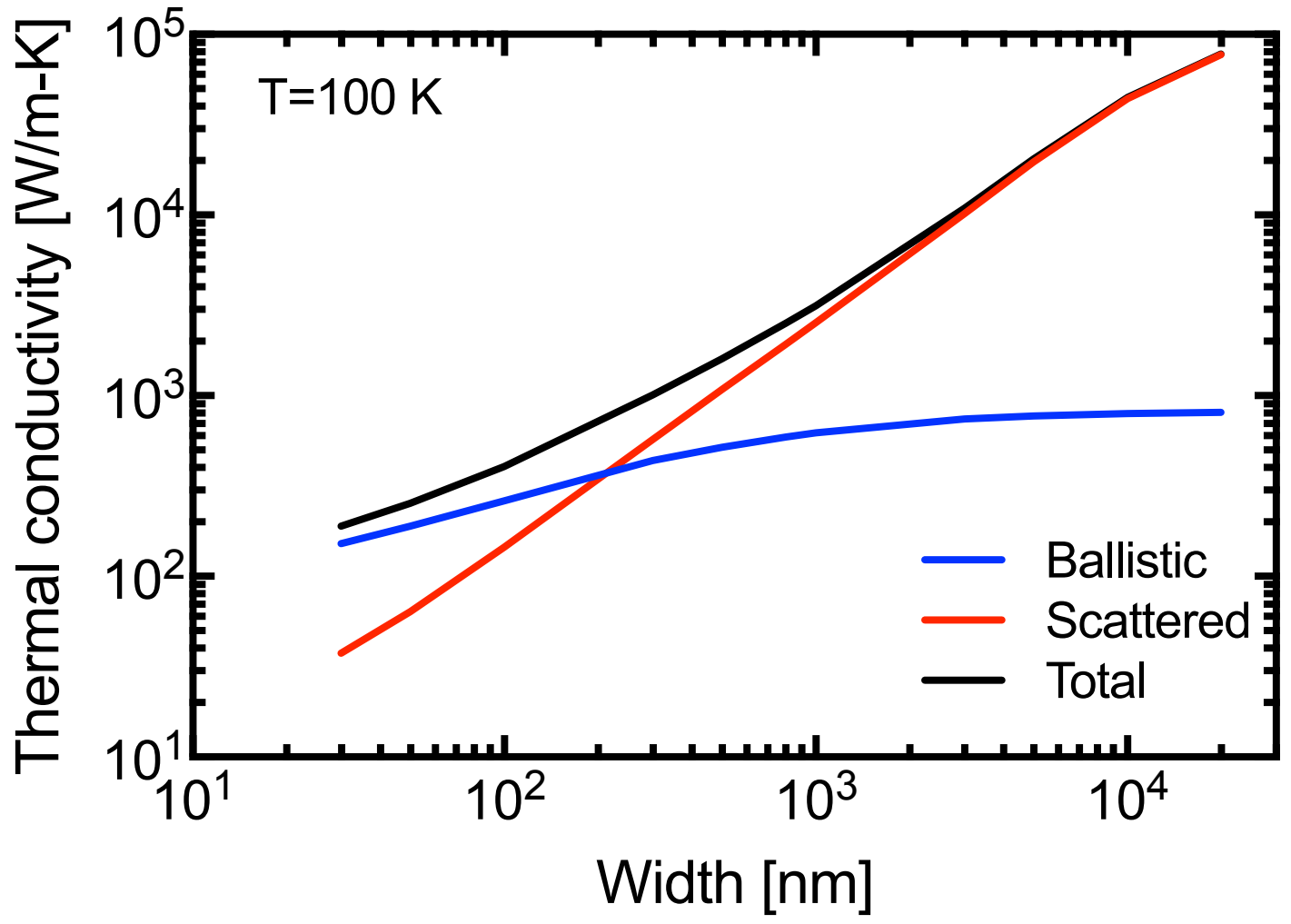

Fig. 2. Ballistic and scattered particles contribution to total thermal conductivity with respect to sample width at 100 K

This decomposition of thermal conductivity clearly distinguishes the ballistic regime from the other two regimes (i.e., the hydrodynamic and diffusive regimes) at small sample width and shows the transition of transport regimes. However, for large sample width, there still exists a need to distinguish between the hydrodynamic and diffusive regimes.

### 3.2. Decomposition of shear stress

A fundamental difference between the hydrodynamic and diffusive regimes lies in the momentum conservation upon a phonon-phonon scattering process. In an ideal hydrodynamic regime where N-scattering is the only internal scattering mechanism, the total phonon momentum is always conserved upon internal phonon scattering; however, the total phonon momentum in an ideal diffusive regime is destroyed upon internal phonon scattering. Therefore, analyzing phonon momentum can serve as a basis to distinguish between the hydrodynamic and diffusive regimes. The $x$-direction momentum ($\Phi_x$) of the phonon system can be defined as the sum of the momentum of all phonon states, i.e.,

$$\Phi_x = \frac{1}{N_i V_{\text{uc}}} \sum_i q_{i,x} n_i \tag{10}$$

Phonon flow is driven by a temperature gradient, creating a net phonon momentum ($\Phi_{\nabla T,x}$). If we assume a steady-state phonon flow, the momentum creations should be balanced with momentum destructions by the direct diffuse boundary scattering without internal phonon scattering (ballistic, $\Phi_{\mathrm{B},x}$), the diffuse boundary scattering combined with N-scattering (hydrodynamic, $\Phi_{\mathrm{H},x}$), or internal U-scattering (diffusive, $\Phi_{\mathrm{D},x}$). The momentum balance can be expressed as

$$\Phi_{\nabla T,x} = \Phi_{\mathrm{B},x} + \Phi_{\mathrm{H},x} + \Phi_{\mathrm{D},x} \tag{11}$$

The momentum gain from a temperature gradient, $\Phi_{\nabla T,x}$, for the computational domain shown in Fig. 1 can be found as

$$\Phi_{\nabla T,x} = \frac{W\delta}{N_i V_{\mathrm{uc}}} \sum_i q_{i,x} v_{i,x} \frac{\partial n_i^0}{\partial T} L\left(-\nabla_x T\right) \tag{12}$$

where $\delta$ is the thickness of graphene, which is assumed as 0.335 nm. The ballistic and hydrodynamic momentum destructions, $\Phi_{\mathrm{B},x}$ and $\Phi_{\mathrm{H},x}$, can be calculated based on the wall shear stress from the ballistic and scattered particles as follows:

$$\Phi_{\mathrm{B},x} = L\delta\left(\tau_{\mathrm{ballistic},y=0} + \tau_{\mathrm{ballistic},y=L}\right) \tag{13}$$

$$\Phi_{\mathrm{H},x} = L\delta\left(\tau_{\mathrm{scattered},y=0} + \tau_{\mathrm{scattered},y=L}\right) \tag{14}$$

Then, the momentum destruction due to internal U-scattering, $\Phi_{\mathrm{D},x}$, is simply calculated as $\Phi_{\nabla T,x} - \Phi_{\mathrm{B},x} - \Phi_{\mathrm{H},x}$ from the momentum balance in Eq. (11).

In Fig. 3, we show the momentum balance as width increases. The black line represents the momentum destruction rate by the wall shear stress of both ballistic and scattered particles $\left(\Phi_{\mathrm{B},x} + \Phi_{\mathrm{H},x}\right)$, calculated from the MC solution of the PBE. The green dashed line shows the momentum gain from a temperature gradient, $\Phi_{\nabla T,x}$, from Eq. (12). When sample width is small and less than 3 μm, the momentum destructions by the wall shear stress are equal to the momentum gains from a temperature gradient, indicating that all the momentum gains from a temperature gradient are dissipated by the wall shear stress and that the internal momentum destruction by U-scattering is negligible. Therefore, for width less than 3 μm, the diffusive regime can be ignored and the actual transport regime is ballistic or hydrodynamic. In order to distinguish between the ballistic and hydrodynamic regimes, we separate the ballistic and scattered particles. When width is smaller than 100 nm, the ballistic contribution is dominant, indicating a strong ballistic regime. For width between 100 nm and 500 nm, the scattered contribution gradually increases, showing a smooth transition from the ballistic to hydrodynamic regimes. For width from 500 nm to 3 μm, the scattered contribution is dominant and the major mechanism of thermal resistance in this width range is the diffuse boundary scattering combined with N-scattering, namely the viscous damping[15]. Again, as $\Phi_{\mathrm{B},x} + \Phi_{\mathrm{H},x}$ is equal to $\Phi_{\nabla T,x}$ and $\Phi_{\mathrm{D},x}$ is negligible, most of the internal phonon scattering processes that particles experience are N-scattering type in this width range. When width is further increased and larger than 3 μm, $\Phi_{\mathrm{B},x} + \Phi_{\mathrm{H},x}$ deviates from $\Phi_{\nabla T,x}$. The difference between $\Phi_{\mathrm{B},x} + \Phi_{\mathrm{H},x}$ and $\Phi_{\nabla T,x}$ represents the momentum destruction by U-scattering according to the momentum balance in Eq. (11). From the width of 3 μm, U-scattering starts to cause thermal resistance and the importance of the diffusive regime gradually increases as width increases.

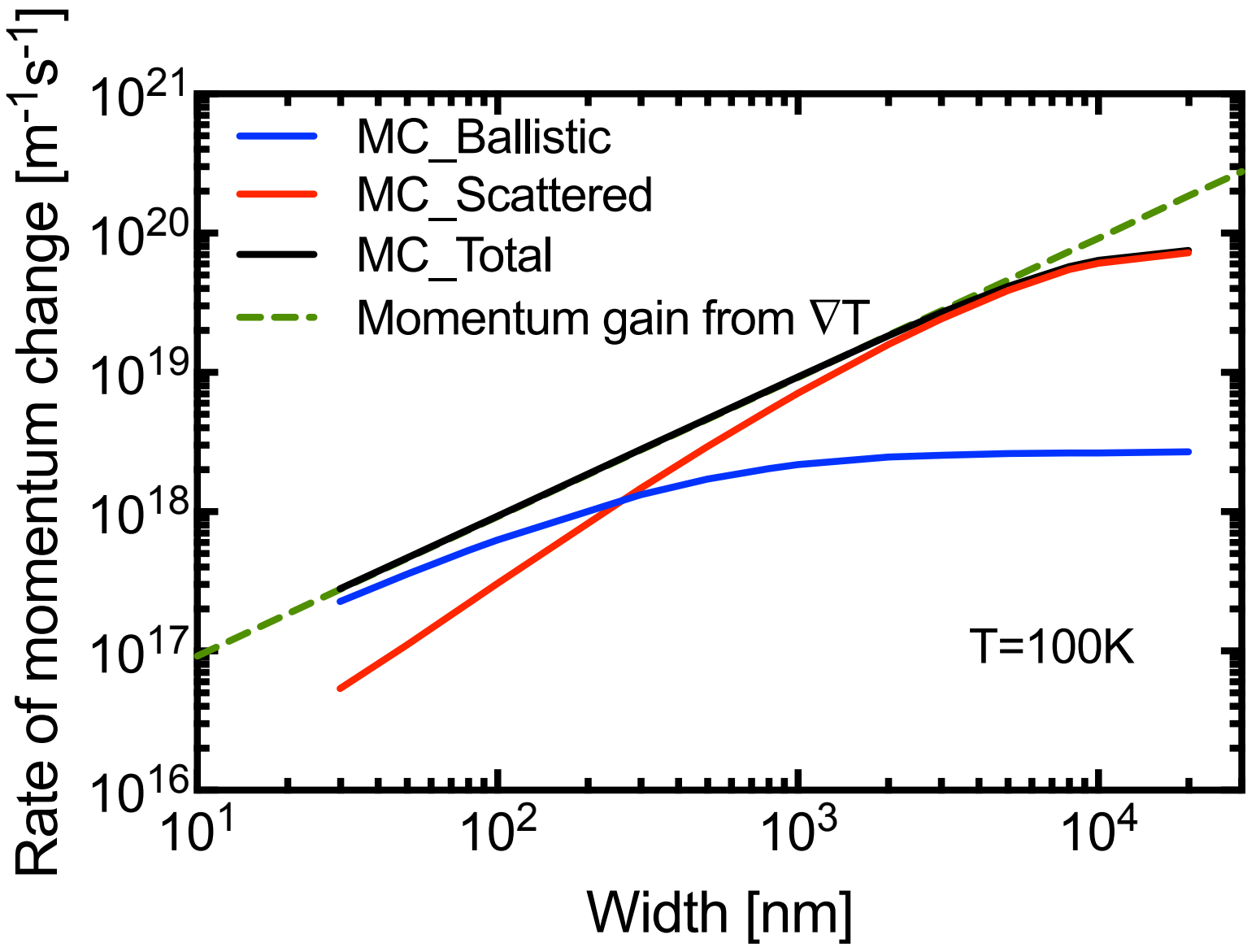

Fig. 3. The momentum balance at 100 K. The blue and red lines represent $\Phi_{\mathrm{B},x}$ and $\Phi_{\mathrm{H},x}$, respectively. The black line is the total momentum destruction rate by the diffuse boundary scattering, $\Phi_{\mathrm{B},x} + \Phi_{\mathrm{H},x}$. The green line is the rate of momentum gain from a temperature gradient, $\Phi_{\nabla T,x}$. The difference between the green and black lines represents $\Phi_{\mathrm{D},x}$.

In Fig. 4, we show the rates of momentum destruction, $\Phi_{\mathrm{B},x}$, $\Phi_{\mathrm{H},x}$, and $\Phi_{\mathrm{D},x}$ normalized by $\Phi_{\nabla T,x}$ at different temperatures. In Fig. 4(a), the ballistic regime is strong for width below 100 nm, while the hydrodynamic regime is significant in a range of width from 300 nm to 10 μm. With sample width increasing, the ballistic-hydrodynamic-diffusive transition is clearly shown in Fig. 4(a). At higher temperatures shown in Fig. 4(b) and 4(c), U-scattering becomes stronger, making the diffusive regime stronger than the 100 K case. The significance of the hydrodynamic regime is thus weakened. The hydrodynamic regime is significant only for a narrow window of sample width or is not strong in the entire range of sample width depending on temperature. For example, at 200 K shown in Fig. 4(b), the hydrodynamic regime accounts for more than 50 % of total momentum destructions in a width range from 300 nm to 500 nm, which is narrower than that at 100 K. When temperature is 300 K as shown in Fig. 4(c), the hydrodynamic regime is not important for any sample width.

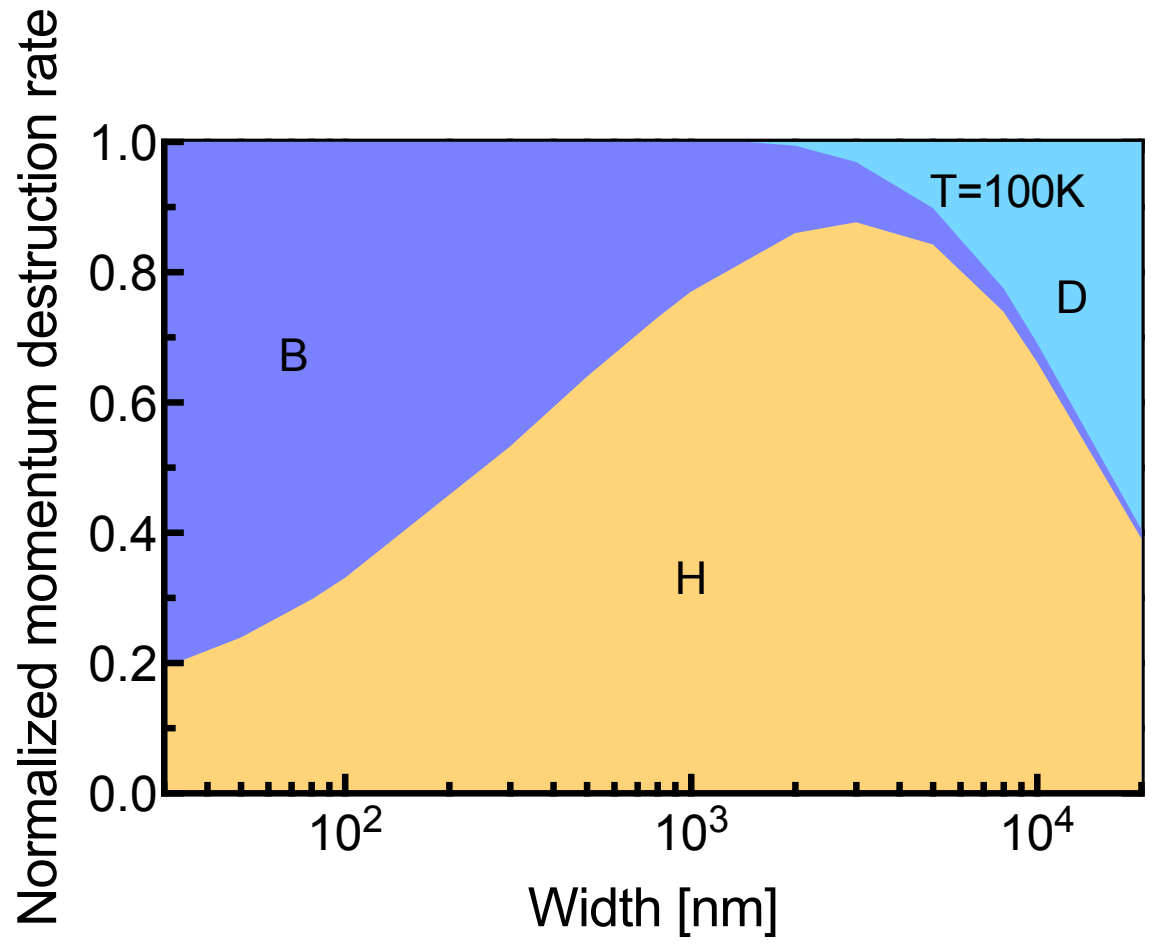

(a)

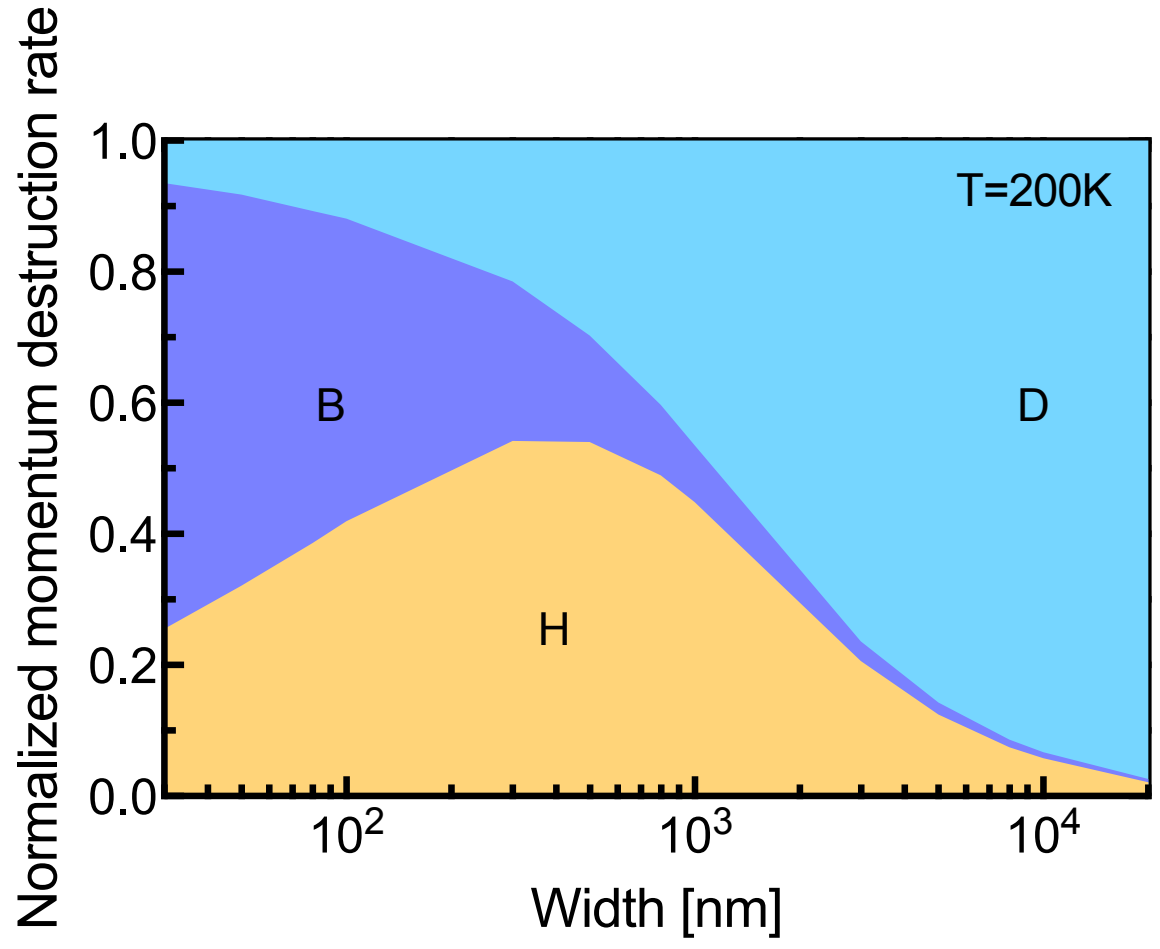

(b)

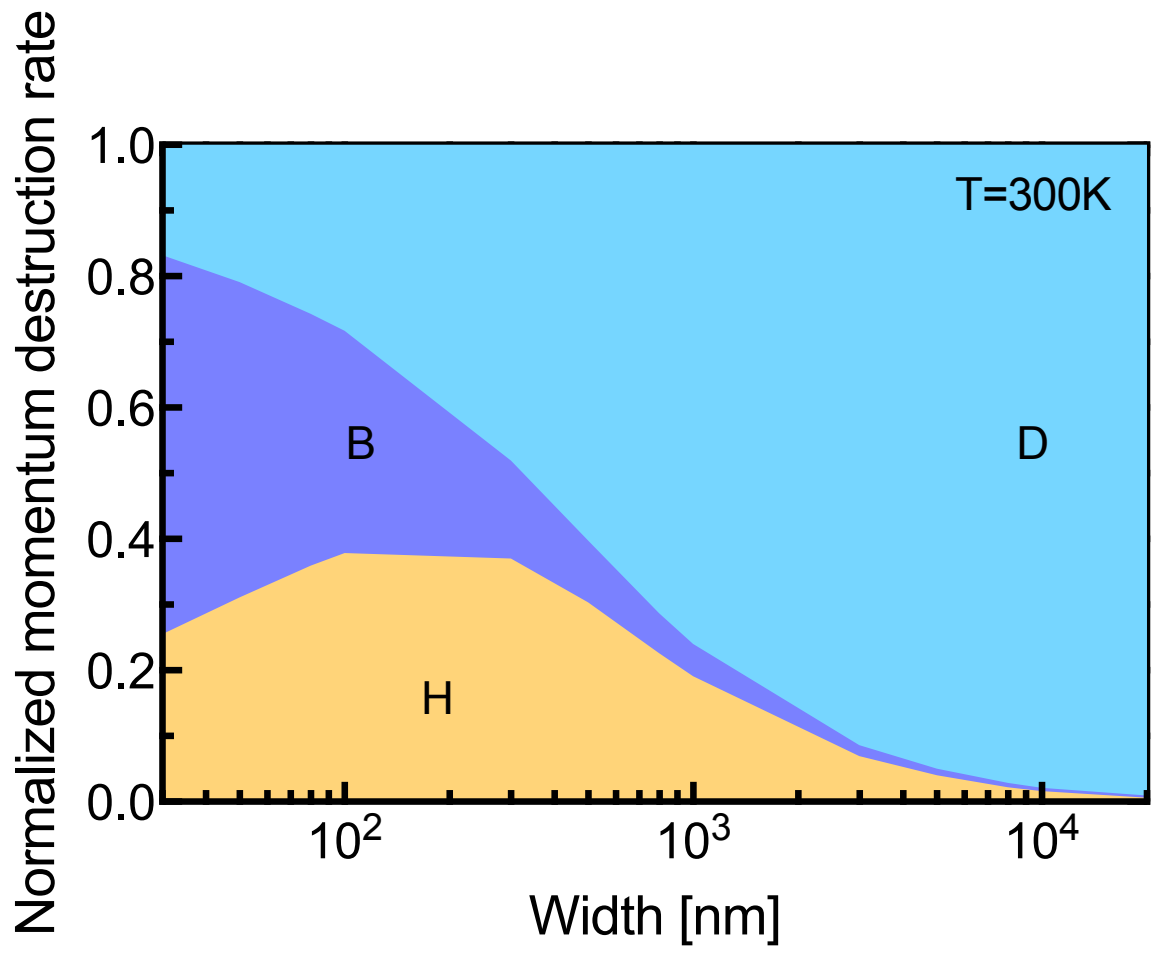

(c)

Fig. 4. Normalized momentum destruction rates at different temperatures, (a) 100 K, (b) 200 K, and (c) 300 K. Dark blue, yellow, and light blue represent the ballistic, hydrodynamic, and diffusive regimes, respectively.

The four-phonon scattering was recently predicted to be important for graphene using the optimized Tersoff potential[6]. The momentum analysis to study the behavior of transport regime crossover can be applicable regardless of the type of phonon scattering and is valid with four-phonon scattering. However, the inclusion of four-phonon scattering would reduce the MFPs of N- and U-scattering from the three-phonon scattering only case, thereby changing the length scale where the crossover occurs shown in Fig. 4.

### 3.3 Phonon Knudsen minimum

The Knudsen minimum is a representative phenomenon of ballistic-to-hydrodynamic transition. Molecular Knudsen minimum was reported by Knudsen[33] around a century ago and had been debated for its existence until the Boltzmann transport equation for molecules was carefully solved in 1960s[34]. Phonon Knudsen minimum was observed at extremely low temperatures[35], [36] and its existence in graphite at much higher temperatures was recently predicted using *ab initio* phonon dispersion and scattering rates[10]. Phonon Knudsen minimum can be found in the dimensionless thermal conductivity defined as

$$\kappa^* = \kappa T_0 / \left( C v_0 W \right) \tag{15}$$

where $C$ is the energy density in phonon hydrodynamics such that $Cu_x$ is the heat flux where $u_x$ is the drift velocity[37]. The $T_0$ is temperature and $v_0$ is an average group velocity. The dimensionless thermal conductivity is defined such that it only depends on Knudsen number and its physical meaning is a space-averaged dimensionless drift velocity at a given dimensionless temperature gradient. A detailed discussion can be found in the Supplementary Information[38] where we provide a semi-analytic solution of the PBE with the Callaway's scattering model for a Debye phonon dispersion.

In Fig. 5, we present the dimensionless thermal conductivity for a Debye phonon dispersion from the semi-analytic solution of the PBE in the Supplementary Information. When there is no U-scattering, i.e., $\tau_{\mathrm{N}} / \tau_{\mathrm{U}} = 0.0$ where $\tau_{\mathrm{N}}^{-1}$ and $\tau_{\mathrm{U}}^{-1}$ represent the N- and U-scattering rates, respectively, the dimensionless thermal conductivity has a minimum point around $Kn = 1$, so called phonon Knudsen minimum. For $\tau_{\mathrm{N}} / \tau_{\mathrm{U}} = 0.1$, the dimensionless thermal conductivity exhibits a slow decrease when inverse Knudsen number is below 5, but then rapidly decreases, leaving no minimum point. As U-scattering rate is further increased, the dimensionless thermal conductivity is converged to the Fuchs-Sondheimer solution[39], [40] which assumes all scattering events destroy the net momentum described by the SMRT approximation.

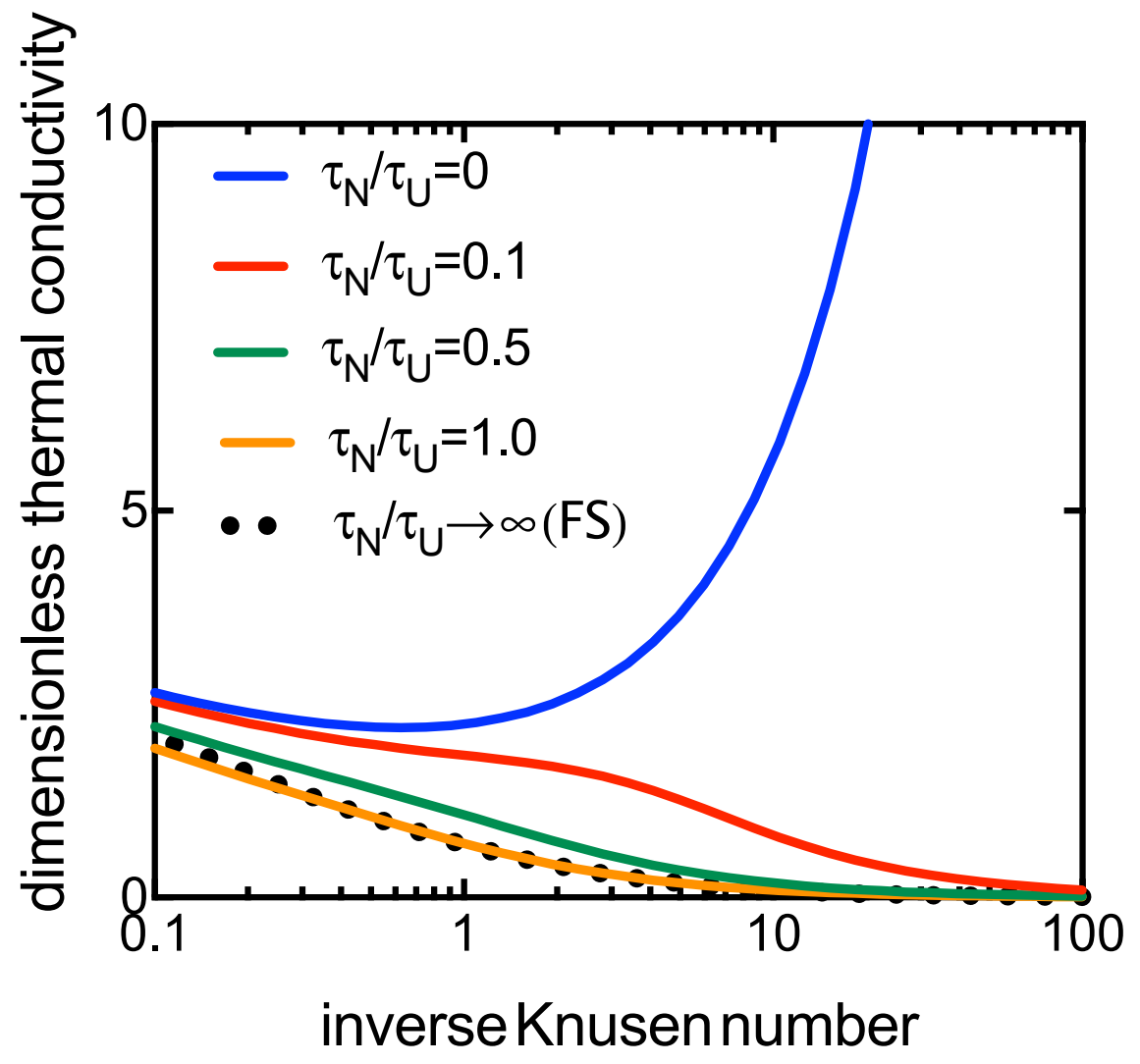

Fig. 5. The dimensionless thermal conductivity ($\kappa^*$) with respect to inverse Knudsen number assuming a Debye phonon dispersion model. The FS refers to the Fuchs-Sondheimer solution of the PBE assuming no N-scattering.

In Fig. 6, we present the thermal conductivity normalized by sample width ( $\kappa / W$ ), similar to the dimensionless thermal conductivity defined in Eq. (15), for suspended graphene at different temperatures. At 100 K, the normalized thermal conductivity exhibits phonon Knudsen minimum when width is around 1 μm, similar to the case without U-scattering in Fig. 5. However, when width becomes larger than 10 μm, the $\kappa / W$ decreases, implying the significant effect of U-scattering on thermal transport. At 200 K, the $\kappa / W$ slowly decreases for width from 100 to 500 nm, but then rapidly decreases without a minimum point, similar to the case of $\tau_N / \tau_U = 0.1$ in Fig. 5. At 300 K, the $\kappa / W$ rapidly decreases for the entire width range.

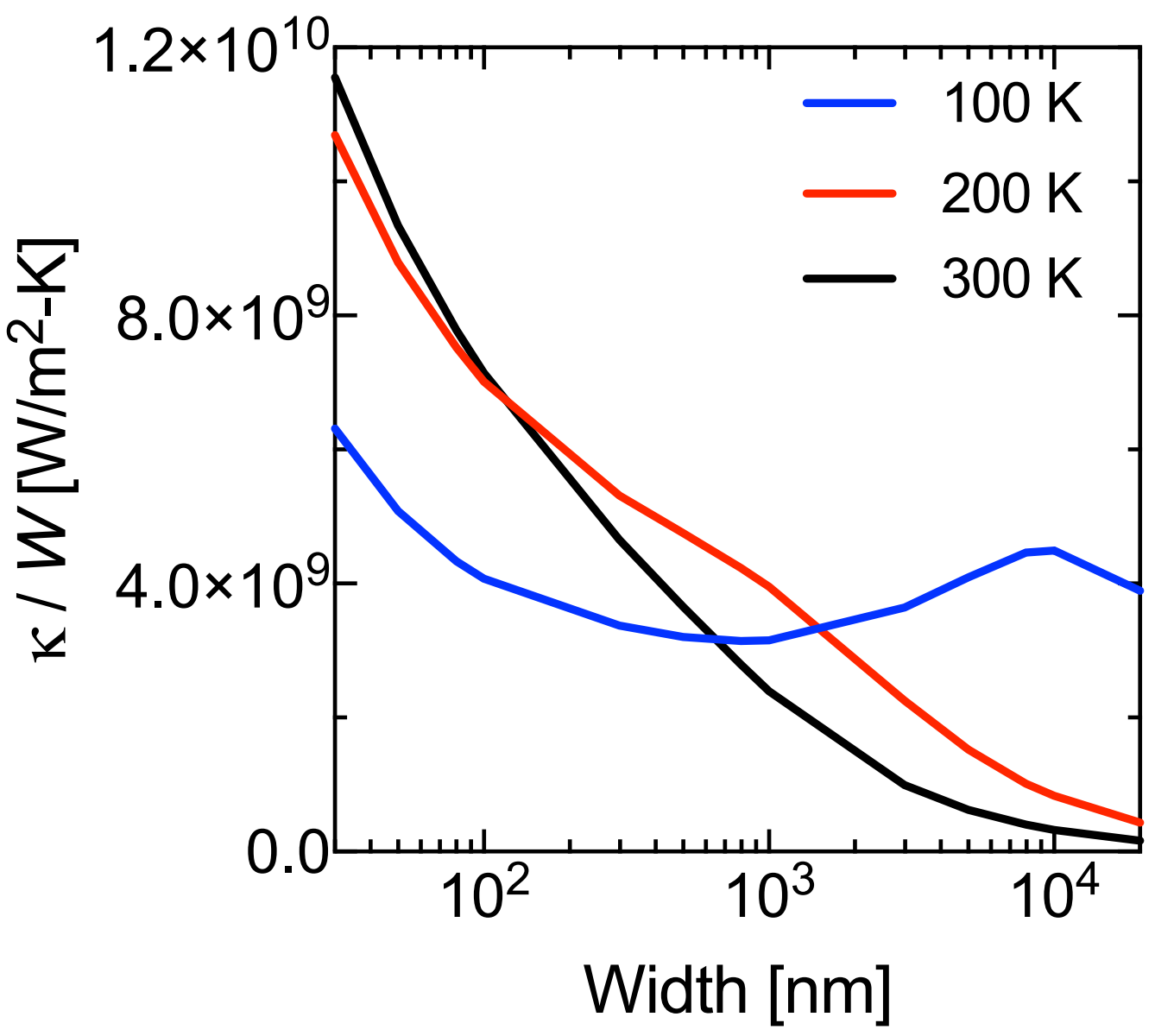

Fig. 6. Thermal conductivity normalized by graphene sample width at 100, 200, and 300 K.

In Fig. 7, we decompose $\kappa / W$ into the ballistic and scattered contributions to further understand Knudsen minimum and the crossover of ballistic, hydrodynamic, and diffusive regimes. At 100 K in Fig. 7(a), the thermal conductivity is mostly from ballistic particles for width below 300 nm, similar to the momentum balance shown in Fig. 4(a). From 300 nm to 10 μm, the scattered contribution is much larger than the ballistic contribution. The scattered contribution of $\kappa / W$ increases with width, and this behavior is particularly significant for width from 1 to 10 μm. The increasing trend of the scattered $\kappa / W$ indicates that the hydrodynamic regime is much more significant than the diffusive regime[15], which also agrees well with the momentum balance in Fig. 4(a). Due to the strong hydrodynamic behavior, the total $\kappa / W$ increases with width, leaving a minimum point at 1 μm. When width is larger than 10 μm, the scattered $\kappa / W$ decreases with width, which agrees well with the momentum balance in Fig. 4(a) showing that the diffusive regime becomes significant at 10 μm.

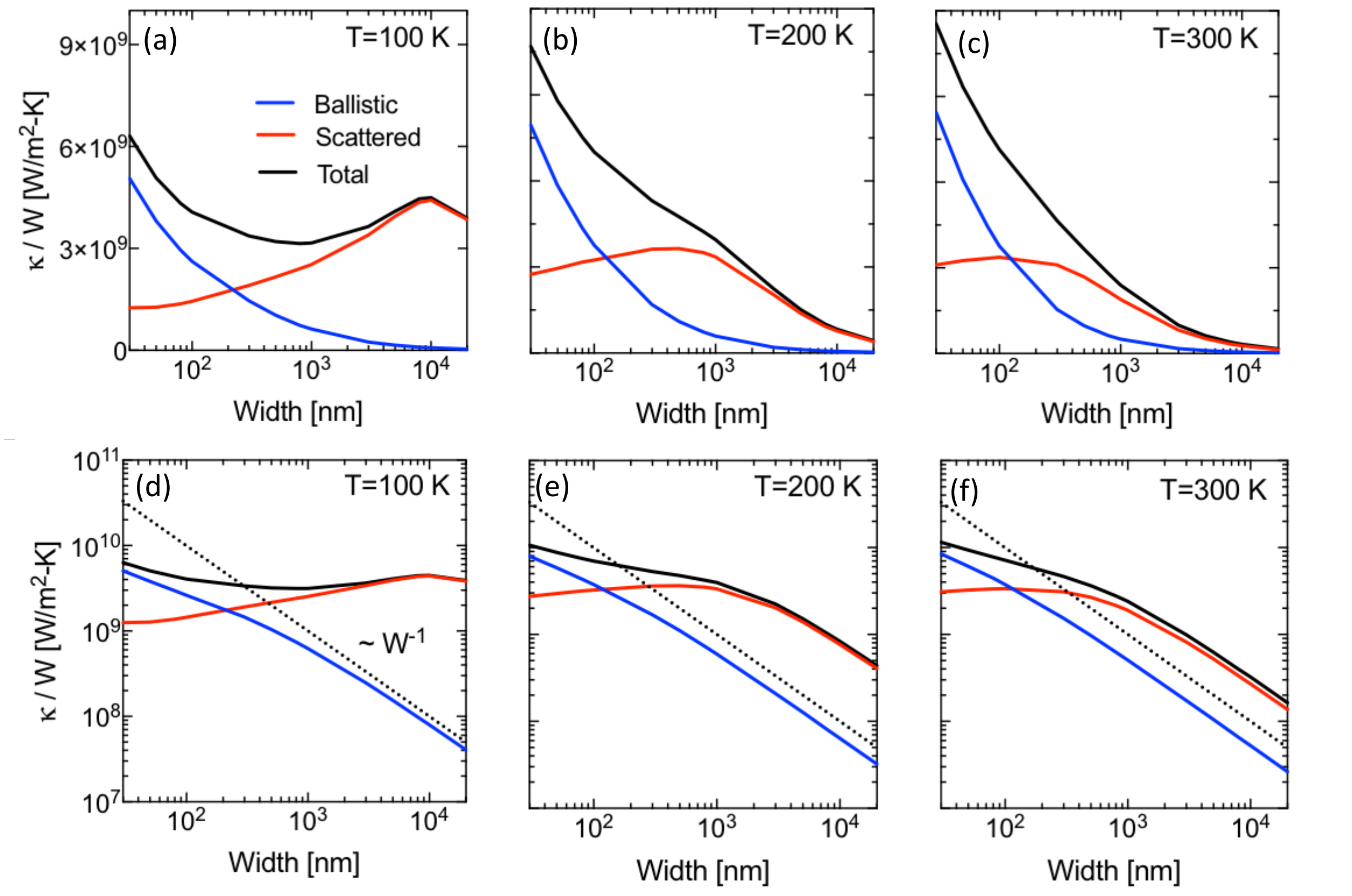

Fig. 7. Normalized thermal conductivity contributions from the ballistic and scattered particles at 100, 200 and 300 K. For (d-f), the $y$-axis is in log scale and $W^{-1}$ dependence is shown as the dotted line for eye-guide.

At 200 K shown in Fig. 7(b), the hydrodynamic behavior exists, but not as significant as the 100 K case. For width from 100 to 500 nm, the scattered $\kappa / W$ increases with width, indicating that the hydrodynamic regime still plays an important role in this width range. This can also be confirmed from the momentum balance in Fig. 4(b) showing significant hydrodynamic regime in the same width range. However, the increasing behavior of the scattered $\kappa / W$ is less significant than the 100 K case, mainly due to larger U-scattering rates. From Fig. 4(b), the momentum destruction by U-scattering already exists for width below 500 nm at 200 K, while it is negligible at 100 K. This results in a slower increase of the scattered $\kappa / W$ and Knudsen minimum cannot be observed at 200 K. For width from 500 nm to 3 μm, the normalized thermal conductivity is mostly due to the scattered particles, but it does not follow $W^{-1}$ dependence which represents constant thermal conductivity and strong diffusive regime (Fig. 7(e)). This shows that the transport is not fully diffusive and the hydrodynamic regime cannot be neglected. When width is larger than 3 μm, the normalized thermal conductivity follows $W^{-1}$, showing the fully diffusive regime where the thermal conductivity is a constant regardless of sample width. When temperature is 300 K (Fig. 7(c) and 7(f)), the hydrodynamic features are further weakened. The normalized thermal conductivity does not show increasing behavior with width for any width larger than 100 nm. When width is larger than 1 μm, the dimensionless thermal conductivity is mostly due to the scattered particles, and it follows $W^{-1}$ trend of the fully diffusive regime.

**4. Conclusions**

We discussed the crossover of the ballistic, hydrodynamic, and diffusive regimes of phonon transport by decomposing the heat flow and wall shear stress into the ballistic and scattered particles contributions. This decomposition framework is applied in our deviational MC simulation of the PBE, where the phonon particles are labeled as scattered or ballistic depending on whether they have experienced internal phonon scattering or not. Using this framework, we compare three different momentum destruction mechanisms and thus measure the significance of all three regimes in a wide range of temperature and sample width. Based on the relative contribution from each transport regime, we could clearly distinguish between the ballistic, hydrodynamic, and diffusive regimes, as well as the transition among these regimes. The characteristic of the transition between the ballistic and hydrodynamic regimes is shown through phonon Knudsen minimum. The proposed decomposition framework would be useful to analyze the significance of all three regimes in high thermal conductivity materials.

## 5. Acknowledgments

This paper is based upon work supported by the National Science Foundation under Grant No. 1705756 and by Central Research Development Funds of University of Pittsburgh. This work used the Extreme Science and Engineering Discovery Environment (XSEDE) Linux cluster at the Pittsburgh Supercomputing Center through allocation TG-CTS180043 and the Linux cluster of the Center for Research Computing at the University of Pittsburgh.